%% file: llr_ref_arxiv.tex
\documentclass[conference]{IEEEtran}
\IEEEoverridecommandlockouts
\usepackage{amsmath}
	\usepackage{amsmath, amsthm, amssymb}
	\usepackage{graphicx}
	\usepackage{epstopdf}
	\usepackage{algpseudocode}% 
	\usepackage{float}
	\usepackage{graphicx}
	\usepackage{amsmath}
	\usepackage{latexsym}
	\usepackage{graphicx}
	\usepackage{bm}
	\usepackage{amssymb}
	\usepackage{array} 
	\usepackage{setspace}
	\usepackage{fancyhdr}
	\usepackage{epstopdf}
	\usepackage{xcolor}
	\usepackage{multicol}
	\usepackage{lipsum}
	\usepackage{multicol,lipsum}
	\usepackage{verbatim}
	\usepackage{calligra}
	\usepackage[english]{babel} % English language/hyphenation
\usepackage{amsmath,amsfonts,amsthm,bm} % Math packages
	\usepackage[scr=boondoxo,
	scrscaled=1.15,]{mathalfa}
	\usepackage{multirow}
	\usepackage{algorithm}
\usepackage{tikz}
\usepackage{pgfplots}
\usepackage{graphicx}
\pgfplotsset{compat=newest}
\usetikzlibrary{shapes,arrows,fit,positioning,calc}
\usetikzlibrary{plotmarks}
\usetikzlibrary{decorations.pathreplacing}
\usetikzlibrary{patterns}
\usetikzlibrary{chains,arrows,shapes,spy}

\usetikzlibrary{automata}
\usepackage{algpseudocode}
\DeclareMathOperator{\tr}{tr}

\def\BibTeX{{\rm B\kern-.05em{\sc i\kern-.025em b}\kern-.08em
    T\kern-.1667em\lower.7ex\hbox{E}\kern-.125emX}}
\begin{document}
\title{\huge Iterative Detection and Decoding Schemes with LLR Refinements in Cell-Free Massive MIMO Networks
 \vspace{-0.15em}
\
}
\pdfinclusioncopyfonts=1
\author{\IEEEauthorblockN{Tonny Ssettumba\IEEEauthorrefmark{1}, Zhichao Shao\IEEEauthorrefmark{2}, Lukas T. N. Landau\IEEEauthorrefmark{1}  and Rodrigo C. de Lamare\IEEEauthorrefmark{1}}	
	\IEEEauthorblockA{\IEEEauthorrefmark{1}CETUC, PUC-Rio, Brazil. Emails:  \{tssettumba, lukas.landau, delamare\}@cetuc.puc-rio.br}
	\IEEEauthorblockA{\IEEEauthorrefmark{2} UESTC, Chengdu 611731, China. Email: zhichao.shao@uestc.edu.cn} \vspace{-0.25em}
 }
\maketitle
\begin{abstract}
In this paper, we propose low-complexity local detectors and log-likelihood ratio (LLR) refinement techniques for a coded cell-free massive multiple input multiple output (CF-mMIMO) systems, where an iterative detection and decoding (IDD) scheme is applied  using parallel interference cancellation (PIC) and access point (AP) selection. In particular, we propose three LLR processing schemes based on the individual processing of the LLRs of each AP, LLR censoring, and a linear combination of LLRs by assuming statistical independence. We derive new closed-form expressions for the local soft minimum mean square error (MMSE)-PIC detector and receive matched filter (RMF). We also examine the system performance as the number of iterations increases.  Simulations assess the performance of the proposed techniques against existing approaches.   
\end{abstract}
\begin{IEEEkeywords}
Iterative scalable cell-free systems, 
receive matched filter,  minimum mean square error filter, parallel interference cancellation. %\vspace{-0.5em}
\end{IEEEkeywords}
\vspace{-0.9em}
\section{Introduction}

Cell-free massive MIMO (CF-mMIMO) enables numerous access points (APs) to serve many single-antenna user equipment (UE) without the need for cell boundaries \cite{rr1,rr3,r9,rr2,itermmsecf,rmmsecf}. Due to the large number of nodes in the network, its topology is almost orthogonal, enhancing coverage, spectrum efficiency, and energy efficiency \cite{rr1}. Additionally, the lack of cell boundaries reduces interference common to conventional cellular systems \cite{rr1}. However, the large number of front-haul connections between APs and the CPU make the network impractical as the number of APs grows. Therefore, scalable CF-mMIMO systems based on dynamic co-operation clustering (DCC) and other selection criteria have been strongly advocated for in previous research \cite{rr2}. These scalable methods perform nearly as well as those that process signals of the entire network while facilitating practical implementations. However, due to multi-user interference (MUI), there are issues with uplink detection for both non-scalable and scalable CF-mMIMO implementations \cite{cesg,rscf,clust&sched,RR13}. Indeed, large-scale MIMO systems \cite{mmimo,wence} require the design of high-quality receivers that can effectively cancel such MUI without increasing complexity and network latency. The key factor in designing an efficient receiver is the reduction of the Euclidean distance between the broadcast signal and the estimated signal. A well-studied CF-mMIMO detector is based on MMSE  with successive interference cancellation (MMSE-SIC) and other nonlinear detectors \cite{mmimo,r2x,jed,r2xy,spa,itic,mfsic,dfcc,mbdf,did,lrcc}. Such an approach, however, is subject to error propagation that increases with each sequential step. Furthermore, the latency for such a detection scheme is quite high and gets worse as the number of UE and antennas in the network  increases, which is inevitable in CF-mMIMO networks \cite{r2x}. When combined with linear and non-linear detection schemes, error correction codes (ECC) such as low-density parity check (LDPC) and Turbo codes can enhance the performance of cellular and CF-mMIMO networks \cite{r10}. The works in \cite{r7} and \cite{r77} consider non-iterative soft detection schemes based on convolutional codes for partial marginalization and distributed expectation detection for CF-mMIMO networks. Because of their cost-effectiveness and properties, LDPC codes have been adopted in the fifth generation (5G) of wireless networks \cite{r10}. Thus,  we adopt LDPC codes in this work.

In the literature, very few works have studied iterative interference cancellation techniques \cite{spa,jed,mfsic,mbdf,aaidd,listmtc,detmtc,msgamp,msgamp2,dynovs} for CF-mMIMO networks, and there is no specific study for both non-scalable and scalable decentralized CF-mMIMO networks. Motivated by our previous works in \cite{r10,r11}, we propose RMF, MMSE, and MMSE-parallel interference cancellation (PIC) techniques \cite{llraps,refidd} for the decentralized CF-mMIMO architecture since they have low complexity and low latency, which are very important for CF-mMIMO networks. %This is due to the fact that 
PIC has been proven to significantly improve IDD techniques because of its robustness in cancelling  MUI in a parallel fashion \cite{r2x}. Moreover, no LLR processing and refinement scheme has been studied for CF-mMIMO networks in the literature. Such LLR refinements can provide significant benefits to CF-mMIMO networks. 

In this work, an iterative CF-mMIMO network with local processing and APs selection (LP-wAPS) is studied. The AP selection is based on the largest large-scale fading (LLSF) coefficients and LLR censoring. Closed-form expressions for the local MMSE-PIC detector are derived considering the interference and AP selection.  
Assuming the absence of prior information on the transmitted code bit at the first iteration, a local linear MMSE detector expression is deducted from the MMSE-PIC expression. Furthermore, the MMSE-PIC detector is compared with the linear MMSE and RMF detectors. Two LLR refinement approaches are then proposed based on LLR censoring and combining and their computational cost is analyzed. These approaches are compared with a standard local processing scheme in which decoding is performed at each AP in terms of average bit error rate (BER) for the entire network. Moreover, the performance of the MMSE-PIC is examined against the number of outer iterations.

The rest of this paper is organized as follows. Section \ref{decentralized} presents the proposed decentralized CF-mMIMO architecture, channel estimation and received signal statistics. The proposed receiver design, computational complexity, and decoding algorithm are presented in Section \ref{Rd}. The proposed LLR processing and refinement schemes are presented in Section \ref{LLRREF}.  Simulation results and discussions are shown
in  Section \ref{Num_Dis}. Section \ref{CO_FD} gives the concluding remarks. 

%\textbf
\textit{Symbol notation}: We use lower/upper bold case symbols to represent vectors/matrices, respectively. The Hermitian transpose operator is denoted by $(\cdot)^{H}$.
% \vspace{-1em}

\section{Proposed System Model} \label{decentralized}

\begin{figure}[htbp]
\centering
\includegraphics[width=7cm]{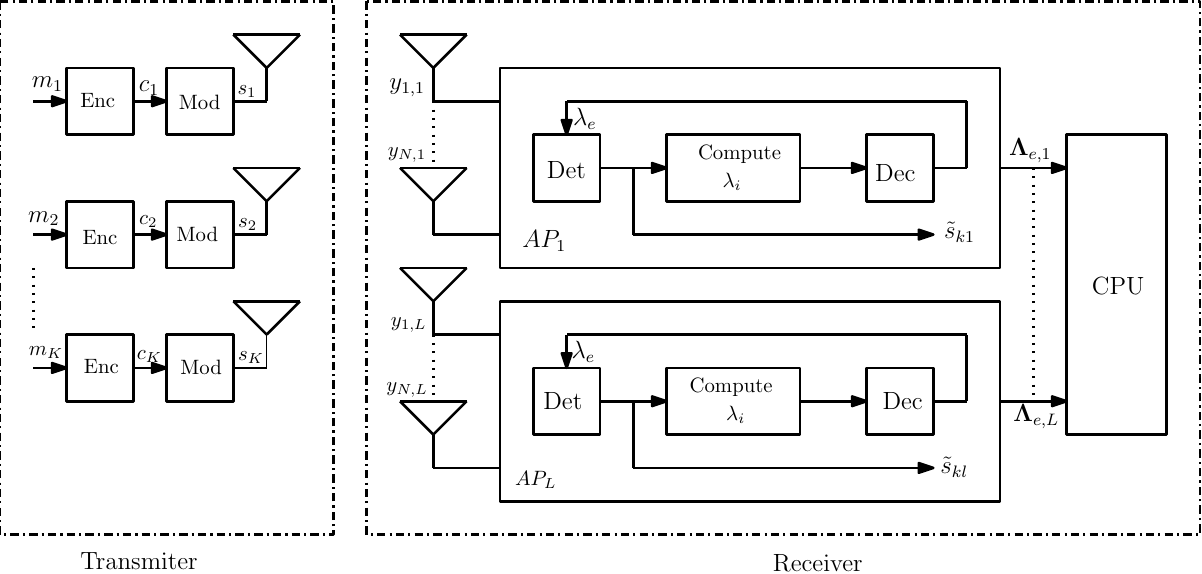}	\caption{Block diagram of the proposed IDD scheme for decentralized CF-mMIMO.}\label{fig1local}
\end{figure}

We consider the uplink of a decentralized CF-mMIMO architecture shown in Figure \ref{fig1local}, comprising of $K$ UEs, an LDPC encoder (Enc) and a modulator (Mod) at the transmitter. The receiver comprises of $L$ APs, each equipped with $N$ receive antennas, a local detector (Det), and a decoder (Dec). The detector can be either RMF, MMSE-PIC, or MMSE that performs the local detection to obtain a detection estimate $\tilde{s}_{kl}$. After detection, local LLRs $\lambda_{i}$ are computed and sent to a  decoder that computes the extrinsic LLRs $\lambda_{e}$. The exchange of these soft beliefs between the detector and decoder improves the performance as the number of iterations increases. The computed LLR streams  $\mathbf{\Lambda}_{e,1}, ..,\mathbf{\Lambda_{e,L}}$ are then sent to the CPU for final processing. The CPU then decides whether to decode the LLRs at each AP  or to censor them based on the AP that provides the most reliable LLRs to a particular UE or to perform a linear combination of the LLRs. 

 \subsection{Uplink Pilot Transmission and Channel Estimation}\label{channel_est}
 
We assume that $\tau_{p}$-length mutually orthogonal pilot sequences $\bm{\psi}_{1}$, ..., $\bm{\psi}_{\tau_{p}}$ with $||\bm{\psi}_{t}||^{2} = \tau_{p}$ are used to estimate the channel. Furthermore, %we assume that 
 $K>\tau_{p}$ is such that more than one UE can be assigned per pilot. The index of UE $k$ that uses the same pilot is denoted as $t_{k} \in \left\{1,..., \tau_{p}\right\}$ with $\vartheta_{k} \subset\left\{1, ..., K\right\}$ as the subset of UEs that use the same pilot as UE $k$ inclusive. 
The complex received signal at the $l$-th AP, \cite{rr1,rr2, r11} $\mathbf{Y}_{l}$, with dimensions  $N\times\tau_{p}$, is given by
 \begin{align}
\mathbf{Y}_{l}=\sum_{j=1}^{K}\sqrt{\eta_{j}}\mathbf{g}_{jl}\bm{\psi}^{T}_{t_{j}}+\mathbf{N}_{l},
 \end{align}
 where $\eta_{j}$ is the transmit power from $j$-th UE, $\mathbf{N}_{l}$ is a receiver noise signal with independent ${\mathcal{N}}_{\mathbb{C}}\sim\left(0, \sigma^{2}\right)$ entries and noise power $\sigma^{2}$, $\mathbf{g}_{jl}\sim\mathcal{N}_{\mathbb{C}}\left(0, \bm{\Omega}_{jl}\right)$, and $\bm{\Omega}_{jl}\in\mathbb{C}^{N\times N}$ is the spatial correlation matrix that describes the channel's spatial properties between the $k$-th UE and the $l$-th AP, $\beta_{k,l}\triangleq \frac{\mathrm{tr}\left(\bm{\Omega}_{jl}\right)}{N}$ is the large-scale (LS) fading coefficient.
%  The first AP %first 
%  correlates the received signal with the associated normalized pilot signal $\bm{\psi}_{t_{k}}/\sqrt{\tau_{p}}$ to 
% $\mathbf{y}_{t_{kl}}\triangleq \frac{1}{\sqrt{\tau_{p}}}\mathbf{Y}_{l}\bm{\psi}^{*}_{t_{k}}\in \mathbb{C}^{N}$ to estimate the channel $\mathbf{g}_{jl}$ given by 
% $\mathbf{y}_{t_{kl}}=\sum_{j\in \vartheta_{k}}\sqrt{\eta_{j}\tau_{p}}\mathbf{g}_{jl}+\mathbf{n}_{t_{kl}}$, 
% where $\mathbf{n}_{t_{kl}}\triangleq \frac{1}{\sqrt{\tau_{p}}}\mathbf{N}_{l}\bm{\psi}^{*}_{t_{k}}\sim\mathcal{N}_{c}\left(0, \sigma^{2}\mathbf{I}_{N}\right)$ is the obtained noise sample after estimation. %Using 
% From \cite{rr1}, the MMSE estimate of $\mathbf{g}_{kl}$ is given by  $\hat{\mathbf{g}}_{kl}=\sqrt{\eta_{k}\tau_{p}}\bm{\Omega}_{kl}\Psi^{-1}_{t_{k}}\mathbf{y}_{t_{kl}}$, where $\Psi_{t_{kl}}=\mathbb{E}\left \{\mathbf{y}_{t_{kl}}\mathbf{y}^{H}_{t_{kl}} \right \}=\sum_{j\in \vartheta_{k}}{\eta_{j}\tau_{p}}\mathbf{\Omega}_{jl}+\mathbf{I}_{N}$ is the received signal vector correlation matrix.
The channel estimate $\hat{\mathbf{g}}_{kl}$ and the estimation error $\tilde{\mathbf{g}}_{kl}=\mathbf{g}_{kl}-\hat{\mathbf{g}}_{kl}$ are independent with distributions $\hat{\mathbf{g}}_{kl}\sim\mathcal{N}_{c}\left(0, \eta_{k}\tau_{p}\bm{\Omega}_{kl}\mathbf{\Psi}^{-1}\bm{\Omega}_{kl}\right)$ and $\tilde{\mathbf{g}}_{kl}\sim\mathcal{N}_{c}\left(0, \mathbf{C}_{kl}\right)$, where the matrix $\mathbf{C}_{kl}$ is given by $\mathbf{C}_{kl}=\mathbb{E}\left \{ \tilde{\mathbf{g}}_{kl} \tilde{\mathbf{g}}^{H}_{kl}\right \}=\bm{\Omega}_{kl}-\eta_{k}\tau_{p}\bm{\Omega}_{kl}\mathbf{\Psi}^{-1}\bm{\Omega}_{kl}$. The received signal at the $l$-th AP is given by
\begin{align}\label{rx_lp}   \mathbf{y}_{l}=\sum_{i=1}^{K}\mathbf{g}_{il}s_{i}+\mathbf{n}_{l}\in\mathbb{C}^{N\times 1}.
\end{align}
The received signal  in \eqref{rx_lp} can be expressed further as
\begin{align}\label{decomposed_sign}  \mathbf{y}_{l}=\hat{\mathbf{g}}_{kl}s_{k}+\sum_{i=1, i\neq k}^{K}\hat{\mathbf{g}}_{il}{s}_{i}+\sum_{m=1}^{K}\tilde{\mathbf{g}}_{ml}s_{m}+\mathbf{n}_{l},
\end{align}
where the first term is the desired signal, the second term is the interference from the other $K-1$ users, the third term denotes the interference due to channel estimation errors and the fourth term denotes the additive white Gaussian noise (AWGN). %denotes 
 \vspace{-0.5em}
\section{ Iterative Local Receiver Design}\label{Rd}
The proposed low-complexity and low-latency local receivers are examined in this section.  The receiver initially computes the symbol mean $\bar{s}_{j}$ to produce soft estimates of the transmitted symbols, which are given by \cite{r10,r11}
\begin{align}\label{expectation_sym}
	\bar{s}_{j}=\sum_{s\in\mathcal{A}}s P(s_{j}=s),
\end{align}
where $\mathcal{A}$  is the set of complex constellations.
The a-priori probability of the extrinsic LLRs is given by
\begin{align}\label{aprior_prob}
	 P(s_{j}=s)=\prod_{l=1}^{M_{c}}\lbrack 1+\exp(-s^{b_{l}}\Lambda_{i}(b_{(j-1)M_{c}+l}))\rbrack^{-1},
\end{align}
where $\Lambda_{i}(b_{i})$ is the extrinsic LLR of the $i$-th  bit calculated by the LDPC decoder in the previous iteration, and $s^{b_{l}}\in (+1,-1)$ denotes  the value of the $l$-th  bit of symbol $s$. The initialization of $\Lambda_{i}(b_{i})$ is obtained from the channel parameters \cite{r10,r11}.
The symbol variance of the $j$-th UE is given by
	\begin{align}\label{variex}
	\sigma^{2}_{j} =\sum_{s\in \mathcal{A}}|s-\bar{s}_{j}|^{2}P(s_{j}=s). 
	\end{align} 
\subsection{Access Point Selection Scheme}
The DCC approach described in \cite{rr2, r11} is taken into account when selecting the APs by forming an APs selection matrix $\mathbf{D}_{kl}$. The APs that provide service to a specific UE are determined by the matrix $\mathbf{D}_{kl}$. By letting $\mathcal {M}_{k}\subset \left\{1,...,L\right\}$ be the subset of APs in service of UE $k$, the matrix $\mathbf{D}_{kl}$ is defined as
\begin{align}
{\mathbf{D}}_{kl}=\left\{\begin{matrix}
\mathbf{I}_{N} & \text{if} & l\in \mathcal {M}_{k}\\ 
 \mathbf{0}_{N}& \text{if}  & l\not\in \mathcal {M}_{k}.
\end{matrix}\right.
\end{align}
Then, the set of UEs that are served by AP $l$ is given by
\begin{align}
    \mathcal{D}_{l}=\biggl\{k:\mathrm{tr}\left(\mathbf{D}_{kl}\right)\geq 1,k\in\left\{1,..,K\right\}\biggr\}.
\end{align} 
% Note that DCC does not alter the statistics of the received signal. It only helps to have a few reliable APs to detect signals.
% The joint APs selection criterion described in \cite{rr2} is used to determine which APs provide service to a specific UE. In this scenario, the UE designates a master AP to coordinate uplink (UL) detection and decoding based on the LLSF. The CPU then establishes a threshold value of the large-scale fading (LSF) coefficient $\beta_{th}$ for non-master APs to serve a particular UE to a certain UE  \cite{rr2}. 
Thus, the estimate of the detected signal of the $k$-th UE at the $l$-th selected AP is given by
\begin{align}\label{mf11}  \tilde{s}_{kl}=\mathbf{w}^{H}_{kl}\mathbf{D}_{kl}\mathbf{y}_{l},
\end{align}
where $\mathbf{w}^{H}_{kl}$ denotes the receive local filter. 
\subsection{Receive Matched Filter}
For the RMF, the weighting vector for the $k$-th UE stream  is given by \cite{rr2}
\begin{align}\label{mf1}  \mathbf{w}_{kl,\text{RMF}}=\mathbf{D}_{kl}\hat{\mathbf{g}}_{kl}. 
\end{align}
By substituting \eqref{mf1} into \eqref{mf11},  the detected signal at the output of a RMF for the $k$-th UE  is given by
\begin{align}\label{mf4}  
&\tilde{s}_{kl}=N\theta s_{k}+\sum_{\gamma=1}^{N}\sum_{i=1, i\neq k}^{K}{\hat{{g}}_{kl,\gamma}}^{*}\hat{{g}}_{il,\gamma}\left[\mathbf{D}_{kl}\right]_{\gamma,\gamma}{s}_{i}\notag\\&+\sum_{\gamma=1}^{N}\sum_{m=1}^{K}{\hat{{g}}_{kl,\gamma}}^{*}\tilde{{g}}_{ml,\gamma}\left[\mathbf{D}_{kl}\right]_{\gamma,\gamma}s_{m}+\sum_{\gamma=1}^{N}{\hat{{g}}_{kl,\gamma}}^{*}\left[\mathbf{D}_{kl}\right]_{\gamma,\gamma}{n}_{l,\gamma}.
\end{align}
The parameter $\theta$ is given by
 $\theta=\frac{1}  {N}\sum_{\gamma=1}^{N}\mid{\hat{{g}}_{kl,\gamma}}\mid^{2}\left[\mathbf{D}_{kl}\right]_{\gamma,\gamma}=\sum_{\gamma=1}^{N}\left( \nu^{2}_{kl,\gamma} +\psi^{2}_{kl,\gamma}\right)\left[\mathbf{D}_{kl}\right]_{\gamma,\gamma},$ 
where $\nu_{kl,\gamma}=\frac{1}{\sqrt{N}} \Re\{\hat{g}_{kl, \gamma}\}$ and $\psi_{kl,\gamma}=\frac{1}{\sqrt{N}} \Im\{\hat{g}_{kl, \gamma}\}$.
%By a
We assume that $\nu_{kl,\gamma}$ and $\psi_{kl,\gamma}$ are Gaussian random variables (RVs) with zero mean and variance $\sigma^{2}_{\nu}=\frac{1}{2N}$. Then, for a selected AP $\theta$ is a Chi-Squared RV with $2N$ degrees of freedom whose mean and variance are given by $\mathbb{E}\left\{ \theta \right\}=2N\sigma^{2}_{\nu}=1 $ and $
\text{Var}\left\{\theta\right\}=4N\sigma^{2}_{\nu}=\frac{1}{N}$, respectively \cite{r2x}. The second order moment of $\theta$ %$\theta=\mathbb{E}\left\{\theta^{2}\right\}$ and 
is mathematically expressed as
\begin{align}\label{mf6}  
  \mathbb{E}\left\{\theta^{2}\right\}= \mathbb{E}\left\{ \theta \right\}+\text{Var}\left\{\theta\right\}=1+\frac{1}{N}. 
\end{align}
From \eqref{mf6}, we can deduct that as the number of antennas $N\gg1$ at each AP,  $\theta^{2} \approx \mathbb{E}\left\{\theta^{2}\right\}\approx 1$. The MUI and the channel estimate error of the RMF are denoted by the second and third terms of \eqref{mf4}. Interference remains constant regardless of the SNR of the channel.  As a result, the RMF's performance for uncoded systems is poor. We propose to enhance its performance with soft RMF and LLR refining techniques. Moreover, a low-latency local iterative interference cancellation strategy based on soft MMSE-PIC is proposed 
in the next subsection.\vspace{-2em}

\subsection{MMSE with Parallel Interference Cancellation}

The local detected symbol estimate of the $k$-th UE  at the  $l$-th AP is obtained by applying an MMSE-PIC filter after subtracting the  expectation of the soft mean values computed in \eqref{expectation_sym}. This is  given by
\begin{align}\label{funAPS1} \tilde{s}_{kl}=&\mathbf{w}^{H}_{kl,\text{PIC}}{\mathbf{D}}_{kl}\hat{\mathbf{g}}_{kl}s_{k}+\mathbf{w}^{H}_{kl,\text{PIC}}{\mathbf{D}}_{kl}\sum_{i=1, i\neq k}^{K}\hat{\mathbf{g}}_{il}\left( {s}_{i}-\bar{s}_{i} \right)\notag\\&+\mathbf{w}^{H}_{kl,\text{PIC}}{\mathbf{D}}_{kl}\sum_{m=1}^{K}\tilde{\mathbf{g}}_{ml}s_{m}+\mathbf{w}^{H}_{kl,\text{PIC}}{\mathbf{D}}_{kl}\mathbf{n}_{l}.
\end{align}
The receive filter $\mathbf{w}^{H}_{kl,\text{PIC}}$ is chosen to minimize the error between the symbol estimate and the transmitted symbol. The optimization problem to obtain $\mathbf{w}^{H}_{kl,\text{PIC}}$ is  formulated as
\begin{align}\label{funp2lp11}
   \mathbf{w}_{kl,\text{PIC}}   =\mathsf{arg}\min_{ \mathbf{w}_{kl,\text{PIC}}}\mathbb{E}\biggl\{||\tilde{s}_{kl}-s_{k}||^{2}\mid \hat{\mathbf{G}}_{l}\biggr\}.
\end{align}
% The objective on the RHS of  \eqref{funp2lp11} is obtained by assuming statistical independence between each term of $\mathbf{y}_{l}$ and ignoring terms that do not depend on $\mathbf{w}^{H}_{kl,PIC}$. After some mathematical and algebraic manipulations, the objective should satisfy the following  relation 
% \begin{align}\label{Objective_fun}
% &\mathbb{E}\biggl\{||\tilde{s}_{kl}-s_{k}||^{2}\mid \hat{\mathbf{G}}_{l}\biggr\}=\mathbf{w}^{H}_{kl, \text{PIC}}{\mathbf{D}}_{kl}\biggl( \rho_{k}\hat{\mathbf{g}}_{kl}\hat{\mathbf{g}}^{H}_{kl}\notag\\&+\sum_{i=1, i\neq k}^{K}\hat{\mathbf{g}}_{il}\mathbb{E}\left\{ \mid {s}_{i}-\bar{s}_{i}\mid^{2} \right\}\hat{\mathbf{g}}^{H}_{il}+\sum_{m=1}^{K}\left ( \mid s_{m}\mid^{2} +\sigma_{m}^{2}\right )\mathbf{C}_{ml}\notag\\&+\sigma^{2}\mathbf{I}_{N} \biggr){\mathbf{D}}^{H}_{kl}\mathbf{w}_{kl, \text{PIC}}-\rho_{k}\mathbf{w}^{H}_{kl,\text{PIC}}{\mathbf{D}}_{kl}\hat{\mathbf{g}}_{kl},
% \end{align}

By differentiating the objective function on the right hand side (R.H.S) of \eqref{funp2lp11} with respect to (w.r.t) $\mathbf{w}^{H}_{kl, \text{PIC}}$ and equating to zero, the optimal local MMSE-PIC filter is given by 
\begin{align}\label{eqlpobjfun1}
&\mathbf{w}_{kl,\text{PIC}}=\rho_{k}\biggl [{\mathbf{D}}_{kl}\biggl(\rho_{k}\hat{\mathbf{g}}_{kl}\hat{\mathbf{g}}^{H}_{kl}+\sum_{i=1, i\neq k}^{K}\hat{\mathbf{g}}_{il}\mathbb{E}\left\{ \mid {s}_{i}-\bar{s}_{i}\mid^{2} \right\}\hat{\mathbf{g}}^{H}_{il}\\&\notag+\sum_{m=1}^{K}\left ( \mid s_{m}\mid^{2} +\sigma_{m}^{2}\right )\mathbf{C}_{ml}+\sigma^{2}\mathbf{I}_{N}\biggr){\mathbf{D}}^{H}_{kl} \biggr ]^{-1}{\mathbf{D}}_{kl}\hat{\mathbf{g}}_{kl}.
 \end{align}
where  $\mathbb{E}\left\{\tilde{\mathbf{g}}_{ml}\tilde{\mathbf{g}}^{H}_{ml}\right\}=\mathbf{ C}_{ml}$, $\mathbb{E}\left\{\mathbf{n}_{l}\mathbf{n}^{H}_{l}\right\}=\sigma^{2}\mathbf{I}_{N}$, $\mathbb{E}\left \{ s_{k}s^{H}_{k} \right \}=\rho_{k}$, $\mathbb{E}\left \{ s_{m}s^{*}_{m} \right \}=\left | s_{m} \right |^{2}+\sigma^{2}_{m}$. Adaptive approaches to computing the MMSE receive filter are also possible \cite{jidf,jiols,jiomimo}. The term $\mathbb{E}\left\{ \mid {s}_{i}-\bar{s}_{i}\mid^{2} \right\}$ denotes the covariance and its values are computed locally at each AP according to  \eqref{variex}.
  
\subsection{Gaussian Approximation of the Soft Local Receive filters}\label{IDD}
% \vspace{-2.1em}

The desired signal, the MUI, the interference from the channel estimation error, and the phase rotated noise constitute the filter's output in \eqref{mf4} and \eqref{funAPS1}. From \cite{r10,r11,r2x}, the filter output of a Gaussian input is Gaussian, which is not necessarily the case for the derived filters. As a result, we assume the filters to be Gaussian, and we derive the mean and variances of the detected signals as follows. The filter output is approximated by  \begin{align}\label{idd1}
  \tilde{s}_{kl} \approx \mu_{kl}s_{k}+\varsigma_{kl}. 
\end{align}
By comparing \eqref{idd1} with \eqref{mf4} and \eqref{funAPS1}, 
$\mu_{kl,\text{RMF}}=N\theta$ and $\mu_{kl,\text{PIC}}=\mathbf{w}^{H}_{kl,\text{PIC}}{\mathbf{D}}_{kl}\hat{\mathbf{g}}_{kl}$. 
The variance $\sigma^{2}_{h}=\mathbb{E}\left\{\mid \tilde{s}_{kl}-\xi_{kl}s_{k}\mid^{2}\right\}=\mathbb{E}\left\{\varsigma_{k}\varsigma^{*}_{k}\right\}$ of $\varsigma_{kl}$ can be obtained using similar steps used in \cite{r7,r10} and is  expressed as 
\begin{align}\label{sdemapcent}
\sigma^{2}_{\text{RMF}}=N\biggl(E_{s}\left ( K-1 \right )+\sum_{m=1}^{K}\left ( \mid s_{m}\mid^{2} +\sigma_{m}^{2}\right )\tr\left\{\mathbf{C}_{ml}\right\}+\sigma^{2}\biggr), 
\end{align} and 
\begin{align}\label{sdemapdeccent}
\sigma^{2}_{\text{PIC}}=&\mathbf{w}^{H}_{kl,\text{PIC}}{\mathbf{D}}_{kl}\biggl ( \sum_{i=1, i\neq k}^{K}\hat{\mathbf{g}}_{il}\mathbb{E}\left\{ \mid {s}_{i}-\bar{s}_{i}\mid^{2} \right\}\hat{\mathbf{g}}^{H}_{il}\notag\\&+\sum_{m=1}^{K}\left ( \mid s_{m}\mid^{2} +\sigma_{m}^{2}\right )\mathbf{C}_{ml}+\sigma^{2}\mathbf{I}_{N}  \biggr ){\mathbf{D}}^{H}_{kl}\mathbf{w}_{kl, \text{PIC}}.
\end{align}
where \eqref{sdemapcent} is based on the assumption that $\mathbf{D}_{kl}=\mathbf{I}_N$.
The extrinsic LLR computed by the detector for the $l$-th bit $l\in\left\{1,2,...,M_{c}\right\}$ of the symbol $s_{k}$ transmitted by the $k$-th user is \cite{r10,r11}
\begin{align}\label{LLR_COMP}
   &\Lambda_{e}\left ( b_{(k-1)M_{c}+l} \right )
   =\notag\\&\frac{\log P\left ( b_{(k-1)M_{c}+l}=+1 |\tilde{s}_{kl}\right)}{\log P\left ( b_{(k-1)M_{c}+l}=-1| \tilde{s}_{kl}\right)}-\frac{\log P\left ( b_{(k-1)M_{c+1}}=+1 \right )}{\log P\left ( b_{(k-1)M_{c+1}}=-1 \right )} \notag\\&
  =\log\frac{\sum _{s\in A^{+1}_{l}}F\left (\tilde{s}_{kl}|s \right )P\left (s \right )}{\sum _{s\in A^{-1}_{l}}F\left (\tilde{s}_{kl}|s \right )P\left (s \right )}-\Lambda_{i}\left ( b_{(k-1)M_{c}+l} \right ),
\end{align}
where the last equality of \eqref{LLR_COMP} follows from Bayes's rule, $A^{+1}_{l}$ is the set of $2^{Mc-1}$ hypotheses for which the $l$-th bit is $+1$. The a-priori probability $P(s)$ is given by \eqref{aprior_prob}. The approximation of the likelihood function $F(\tilde{s}_{kl}|s)$ is given by  \cite{r10,r11} \begin{align}\label{llfn}
        F\left ( \tilde{s}_{kl}|s \right )\simeq\frac{1}{\pi\sigma^{2}_{h}}\exp\left (-\frac{1}{\sigma^{2}_{h}} |\tilde{s}_{kl}-\mu_{kl,h}s|^{2} \right ).
    \end{align}
    \subsection{Computational Complexity}
Table \ref{CC} presents the number of required flops per local detector. It can be observed  that the computational complexity of the RMF and MMSE based detectors are of the order $\mathcal{O}(K^{2}L)$ and $\mathcal{O}(N^{2}LK)$, respectively, where $\mathcal{O}(\cdot)$ is the Big O notation.
\begin{table}[h!]
\renewcommand*{\arraystretch}{1.5}
\begin{footnotesize}
\caption{Computational complexity per detector.}
\vspace{-1em}
\begin{center}
\begin{tabular}{|p{1.8cm}|p{5cm}|}
\hline
\textbf{Local Detector} & \textbf{{Number of multiplications or flops }} \\
\hline
RMF & $2K^{2}L+4KNL+4KLM_{c}2^{M_{c}}+4KL2^{M_{c}}$  \\
\hline
MMSE & $2N^{2}LK+2K^{2}NL+8KNL+4KL2^{M_{c}}+2M_{c}KL2^{M_{c}}+KL$ \\
\hline
MMSE-PIC & $4N^{2}LK+3K^{2}NL+8KNL+9KL2^{M_{c}}+4M_{c}KL2^{M_{c}}+KL$   \\
\hline
\end{tabular}
\label{CC}
\end{center}
\end{footnotesize}
\end{table}
% \vspace{-0.95em}
\subsection{Decoding Algorithm}
The proposed detectors and decoders exchange soft beliefs in an iterative fashion. The tangent function degrades the conventional sum-product algorithm (SPA) performance in the error floor region \cite{r11}. For this reason, we use the Box-plus SPA as in our previous papers \cite{r10,r11} since it produces less complex approximations.
% The LLR sent from the check node $(CN)_{J}$ to the variable node $(VN)_{i}$ is computed as 
% \begin{align}
% \Lambda_{j\longrightarrow i}=\boxplus_{i^{'}\in N(j)\diagdown i} \Lambda_{i^{'\longrightarrow j}},
% \end{align}
% where $\boxplus$ denotes the pairwise ``box-plus" operator given by 
% \begin{align}
% \Lambda_{1}\boxplus \Lambda_{2}=&\log\left ( \frac{1+e^{\Lambda_{1}+\Lambda_{2}}}{e^{\Lambda_{1}}+e^{\Lambda_{2}}} \right ),\\\notag
%   = \hspace{.1cm}& \mathrm{sign}(\Lambda_{1})\mathrm{sign}(\Lambda_{2})\min(\left | \Lambda_{1} \right |,\left | \Lambda_{2} \right |)\\\notag&+\log\left ( 1+e^{-\left |\Lambda_{1}+\Lambda_{2}  \right |} \right )-\log\left (1+e^{-\left |\Lambda_{1}-\Lambda_{2}  \right |}  \right ).
% \end{align}
% The LLR from $VN_{i}$ to $CN_{j}$ is given by
% \begin{align}   \Lambda_{i\longrightarrow j}=\Lambda_{i}+\sum_{j^{'}\in {N(i)\backslash j}}\Lambda_{j^{'}\longrightarrow i}, 
% \end{align}
% where the parameter $\Lambda_{i}$ denotes the LLR at $VN_{i}$, ${j^{'}\in N(i)\backslash j}$ %denotes 
% {means that }all $CNs$ are connected to $VN_{i}$ except $CN_{j}$.

\section{Proposed LLR Processing and Refinement Strategies}\label{LLRREF}
In this section, we present the proposed LLR refinement strategies and discuss how they work.
\subsection{Standard LLR Processing }

In this scheme, BER is computed based on LLRs for each AP. Then, the average BER is obtained for the entire network. This approach adversely affects the network performance because some APs have unreliable LLRs for articular UEs. Thus, we propose two strategies based on LLR censoring and combining to improve the performance, as discussed below.

\subsection{LLR Censoring}
We present an LLR censoring technique that helps reduce the redundant processing of LLRs at the CPU. The independent streams of LLRs are sent from the APs to the CPU with dimension $KC_{\text{leng}}L$, where $C_{\text{leng}}$ is the codeword length. At each AP, the mean absolute value of the LLRs is computed as
\begin{align}
\mu_{\Lambda_{kl,e}}=\frac{1}{C_{\text{leng}}}\sum_{c=1}^{C_{\text{leng}}}\rvert\Lambda_{l,e}\rvert.
\end{align}
Based on $\mu_{\Lambda_{kl,e}}$, the UE is decoded at the AP where this parameter is highest and the other LLRs are discarded. This is done for all APs and a new matrix with the censored LLRs of dimensions $KC_{\text{leng}}$ is formed to perform the final decoding.

 \subsection{LLR Combining} 
In this refinement strategy, we compute the sum of the multiple streams of LLRs coming from the different APs by assuming statistical independence. The combined LLRs are mathematically given by
\begin{align}\label{LLR_AVG}
\Lambda_{\rm avg,e} \left( b_{(k-1)M_{c}+l} \right) = %\frac{1}{L} 
\sum_{l=1}^{L} \Lambda_{l,e} \left( b_{(k-1)M_{c}+l} \right).
\end{align}
The idea of combining multiple streams of LLRs creates some diversity benefits from the LLRs which yields performance benefits.
The mean of the refined LLRs is given by
\begin{align}\label{LLR_mean}
E[\Lambda_{\rm avg,e} \left( b_{(k-1)M_{c}+l} \right)] & =  \sum_{l=1}^{L} E[\Lambda_{l,e} \left( b_{(k-1)M_{c}+l} \right)]\\
& = \mu_{\Lambda_{\rm avg,e} },
\end{align}
and the variance of the refined LLRs is %given by
\begin{equation}
\begin{split}
\sigma^2_{\Lambda_{\rm avg,e}} & = E[ \left(\Lambda_{\rm avg,e} \left( b_{(k-1)M_{c}+l} \right) - \mu_{\Lambda_{\rm avg,e}} \right)^2 ] \\
& = \frac{1}{L} \sum_{l=1}^{L} \Big((\Lambda_{\rm avg,e} \left( b_{(k-1)M_{c}+l} \right)^2 \\ & \quad -  \sum_{n=1}^{L} \Lambda_{n,e} \left( b_{(k-1)M_{c}+l}) \right).
\end{split}
\end{equation}
This suggests that the refinement benefits come from improving the quality of the LLRs through their variance reduction that shifts the LLRs with small values away from the origin. 

\section{Simulation Results}
\label{Num_Dis}

The BER performance for CF-mMIMO settings is presented in this section. Due to LS fading coefficients, the CF-mMIMO channel exhibits high PL values. Thus, the SNR at the $l$-th AP is expressed as
\begin{align}
\mathbb{SNR}_{l}=\frac{\tr(\mathbf{G}_{l}~\mathrm{diag}\left(\bm{\rho}\right)\mathbf{G}^{H}_{l})}{\sigma_{w}^{2} N K}.
\end{align}
Unless stated otherwise, the simulation parameters are varied as follows.
We consider a cell-free environment with a square of dimensions $D\times D$, where $D=1$ km. The APs are deployed $10~$m above the UE. The bandwidth is 20 MHz, $d_{th}=0.38$, $\tau_{u}=190$, $\tau_{p}=10$, $\tau_{c}=200$, $\eta_{k}=100$ mW. The spatial correlation matrices $\bm{\Omega}_{jl}$ are assumed to be locally available at the APs \cite{rr2}. We use LDPC codes \cite{pegbf,vfap,memd}
with codeword length $C_{\text{leng}}=256$ bits, $M=128$ parity check bits and $C_{\text{leng}}-M$ message bits. The threshold for the non-master AP is set to $\beta_{\text{th}}=-40~\mathrm{dB}$ and the code rate is $R=\frac{1}{2}$. The maximum number of inner iterations  (decoder iterations) is set to $10$. The signal power $\rho=1$ W and the simulations are run for $10^{4}$ channel realizations. The modulation scheme is QPSK. The LS fading coefficients are obtained according to the Third Generation Partnership Project (3GPP) Urban Microcell model in \cite{rr2} given by $
\beta_{k,l}\left[\mathbf{\mathrm{dB}}\right]=-30.5-36.7\log_{10}\biggl(\frac{d_{kl}}{1 m}\biggr)+\Upsilon_{kl},
$ where $d_{kl}$ is the distance between the $k$-th UE and $l$-th AP,  $\Upsilon_{kl}\sim\mathcal{N}\left(0, 4^{2}\right)$ is the shadow fading. 

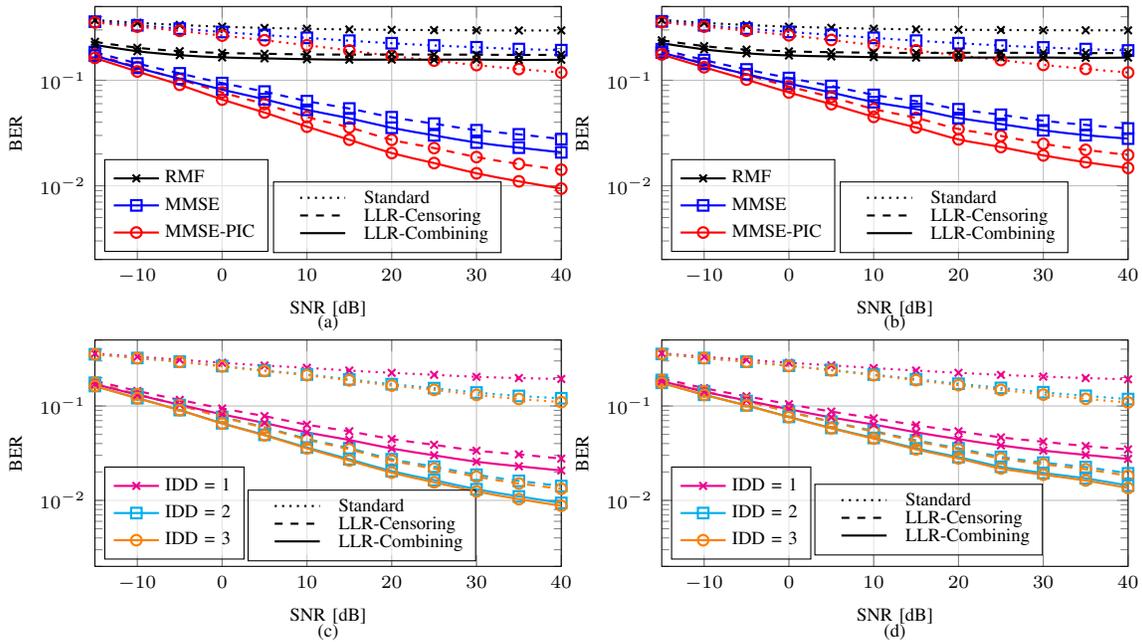
\begin{figure*}[!htbp]
\centering
 \input{BER_cf_vs_col}
 \input{IDDITERVARY}
\caption{BER versus SNR while comparing the studied detectors and LLR refinement strategies  for $L=4$, $N=4$, $K=4$: (a) all APs (Full-Network),  $\text{IDD}=2$ (b) LP-wAPS (Scalable),  $\text{IDD}=2$ (c) Full-Network  for  MMSE-PIC, (d) LP-wAPS (Scalable) with MMSE-PIC.}
 \label{figJ1} 
 \end{figure*}
 
The BER versus SNR characteristic for the three LLR processing schemes is presented in Figures \ref{figJ1}(a) and \ref{figJ1}(b) 
for all APs and  scalable networks, respectively. The MMSE-PIC detector achieves the lowest BER, followed by the MMSE and lastly the RMF. The LLR combining and censoring schemes achieve the best performance, while individual decoding at each AP achieves the worst performance. This is why the LLR refinements improve the network's performance. The superior performance of the PIC-based detector is attributed to its ability to cancel MUI in a robust manner. On the other hand, RMF has poor performance but does not require any matrix inversions, which reduces the complexity. Furthermore, %the 
LLR censoring helps to mitigate redundant processing at the CPU. However, because the CPU must constantly search through all APs to identify the one with the most reliable LLRs for a specific UE, which increases the hardware complexity of the receiver. The superior performance of LLR combining is due to its ability to achieve LLR diversity and shift LLRs with poor beliefs away from the origin. Figures \ref{figJ1}(c) and \ref{figJ1}(d) are, respectively,  for all APs and scalable networks with  
 the MMSE-PIC detector. Varying iterations from one to two provides a significant reduction in BER for LLR processing schemes but a marginal reduction after the second iteration. The exchange of soft beliefs between decoders and detectors improves performance as it is useful in the cancellation step.
 
\section{Concluding Remarks}\label{CO_FD}
In this paper, novel LLR refinement strategies for IDD schemes in CF-mMIMO networks have been proposed. Low-complexity and low-latency local receive filters based on RMF and MMSE-PIC have been studied for the decentralized CF-mMIMO network.
Finally, the LLR refinement strategies based on LLR censoring and LLR combining were devised to improve the performance of CF-mMIMO systems.

\end{document}

%% file: BER_cf_vs_col.tex
\definecolor{mycolor1}{rgb}{0.00000,1.00000,1.00000}%
\definecolor{mycolor2}{rgb}{1.00000,0.00000,1.00000}%
\definecolor{mycolor3}{rgb}{0.83,0.69,0.22}%

\pgfplotsset{every axis label/.append style={font=\scriptsize
},
every tick label/.append style={font=\scriptsize
}
}

\begin{tikzpicture}[font=\scriptsize
] 
\begin{axis}[%
name=IF1,
%ymode=log,
width  = 0.7\columnwidth,%5.63489583333333in,
%height = 0.3\columnwidth,%4.16838541666667in,
height = 0.38\columnwidth,%4.16838541666667in,
scale only axis,
xmin  = -15,
xmax  = 40,
xlabel= {SNR [dB]},
xmajorgrids,
ymin=0.002,
ymax=0.5,
ymode=log,
ylabel={BER},
ymajorgrids,
legend entries={RMF,
					MMSE,
				MMSE-PIC,	
				%MMSE based comparator network,				
				},
legend style={fill=white, fill opacity=0.6, draw opacity=1,
text opacity =1,at={(0.02,0.03)}, anchor= south west,draw=black,fill=white,legend cell align=left,font=\scriptsize}
]

\addlegendimage{smooth,color=black,solid, thick, mark=x,
y filter/.code={\pgfmathparse{\pgfmathresult-0}\pgfmathresult}}
\addlegendimage{smooth,color=blue,solid, thick, mark=square,
y filter/.code={\pgfmathparse{\pgfmathresult-0}\pgfmathresult}}
\addlegendimage{smooth,color=red,solid, thick, mark=o,
y filter/.code={\pgfmathparse{\pgfmathresult-0}\pgfmathresult}}
\addlegendimage{smooth,color=magenta,solid, thick, mark=o,
y filter/.code={\pgfmathparse{\pgfmathresult-0}\pgfmathresult}}
\addlegendimage{smooth,color=green,solid, thick, mark=diamond,
y filter/.code={\pgfmathparse{\pgfmathresult-0}\pgfmathresult}}
% Add a (b) below x-axis
\pgfplotsset{
    every axis/.append style={
        extra description/.code={
            \node at (0.5,-0.25) {(a)};
        },
    },
}

% \draw (31,0.18) ellipse (0.2cm and 0.6cm);
% \draw[dspconn]    (30,0.25) -- (20,0.3) ;
% \draw (12,0.4) node [anchor=north west][inner sep=0.75pt]  [font=\footnotesize]  {\text{All-APs}};

% \draw (35,0.05) ellipse (0.2cm and 0.9cm);
% \draw[dspconn]   (30,0.04) -- (10,0.055) ;
% \draw (4,0.07) node [anchor=north west][inner sep=0.75pt]  [font=\footnotesize]  {\text{APs-Sel}};

%% IMPERFECT CSI
\addplot+[smooth,color=black,dotted,thick, every mark/.append style={solid} ,mark=x,
y filter/.code={\pgfmathparse{\pgfmathresult-0}\pgfmathresult}]
  table[row sep=crcr]{%
-15	0.372411303710938	\\
-10	0.350522412109375\\
-5	0.334654516601563	\\
0	0.321956616210938\\
5	0.313611401367188	\\
10	0.307889794921875\\
15	0.303165136718750	\\
20	0.300699584960938\\
25	0.298820214843750	\\
30	0.297485083007813	\\
35	0.296502880859375	\\
40	0.296896142578125	\\
%25	0	\\
};
%\addlegendentry{MMSE col}

\addplot+[smooth,color=blue,dotted, thick, every mark/.append style={solid} ,mark=square,
y filter/.code={\pgfmathparse{\pgfmathresult-0}\pgfmathresult}]
  table[row sep=crcr]{%
-15	0.359346777343750	\\
-10	0.331952075195313\\
-5	0.308786645507813	\\
0	0.286850000000000\\
5	0.268600756835938	\\
10	0.252477050781250\\
15	0.237208813476563	\\
20	0.224407348632813\\
25	0.213825292968750	\\
30	0.205531347656250\\
35	0.197682348632813	\\
40	0.192936669921875	\\
};

%\addlegendentry{SIC col}

\addplot+[smooth,color=red,dotted, thick, every mark/.append style={solid} ,mark=o,
y filter/.code={\pgfmathparse{\pgfmathresult-0}\pgfmathresult}]
  table[row sep=crcr]{%
-15	0.352959521484375	\\
-10	0.322147924804688\\
-5	0.294060180664063	\\
0	0.265165332031250\\
5	0.239342700195313	\\
10	0.214815185546875	\\
15	0.191652270507813	\\
20	0.170120361328125\\
25	0.153654736328125	\\
30	0.139308203125000	\\
35	0.127775048828125\\
40	0.118456445312500	\\
};
\addplot+[smooth,color=black,dashed,thick, every mark/.append style={solid} ,mark=x,
y filter/.code={\pgfmathparse{\pgfmathresult-0}\pgfmathresult}]
  table[row sep=crcr]{%
-15	0.231298632812500	\\
-10	0.203281250000000\\
-5	0.188772460937500	\\
0	0.181527441406250\\
5	0.178310742187500	\\
10	0.175939355468750\\
15	0.175052148437500	\\
20	0.175574316406250\\
25	0.175154199218750	\\
30	0.174819921875000	\\
35	0.173862500000000\\
40	0.174307226562500	\\
%25	0	\\
};
%\addlegendentry{MMSE col}

\addplot+[smooth,color=blue,dashed, thick, every mark/.append style={solid} ,mark=square,
y filter/.code={\pgfmathparse{\pgfmathresult-0}\pgfmathresult}]
  table[row sep=crcr]{%
-15	0.184833886718750	\\
-10	0.144235839843750\\
-5	0.116271777343750	\\
0	0.094086914062500\\
5	0.078166015625000	\\
10	0.063386718750000\\
15	0.053987597656250	\\
20	0.044621484375000\\
25	0.038811425781250	\\
30	0.033607812500000\\
35	0.030712109375000	\\
40	0.027699414062500	\\
};

%\addlegendentry{SIC col}

\addplot+[smooth,color=red,dashed, thick, every mark/.append style={solid} ,mark=o,
y filter/.code={\pgfmathparse{\pgfmathresult-0}\pgfmathresult}]
  table[row sep=crcr]{%
-15	0.176170507812500	\\
-10	0.133085449218750\\
-5	0.102417675781250	\\
0	0.076673535156250\\
5	0.059537011718750	\\
10	0.044890234375000	\\
15	0.035723632812500	\\
20	0.027275976562500\\
25	0.022764843750000	\\
30	0.018699218750000\\
35	0.016093945312500	\\
40	0.014187597656250	\\
};
\addplot+[smooth,color=black,solid,thick, every mark/.append style={solid} ,mark=x,
y filter/.code={\pgfmathparse{\pgfmathresult-0}\pgfmathresult}]
  table[row sep=crcr]{%
-15	0.216437011718750	\\
-10	0.188419042968750\\
-5	0.172975781250000	\\
0	0.164773339843750\\
5	0.161342285156250	\\
10	0.158352050781250\\
15	0.156782031250000	\\
20	0.156989843750000\\
25	0.156662597656250	\\
30	0.157039062500000	\\
35	0.155757812500000	\\
40	0.156557226562500	\\
%25	0	\\
};
%\addlegendentry{MMSE col}

\addplot+[smooth,color=blue,solid, thick, every mark/.append style={solid} ,mark=square,
y filter/.code={\pgfmathparse{\pgfmathresult-0}\pgfmathresult}]
  table[row sep=crcr]{%
-15	0.171002832031250	\\
-10	0.131316210937500\\
-5	0.103304394531250	\\
0	0.081508984375000\\
5	0.066121191406250	\\
10	0.052459960937500\\
15	0.043685644531250	\\
20	0.035437402343750\\
25	0.030042871093750	\\
30	0.025676269531250\\
35	0.022973632812500	\\
40	0.020736621093750	\\
};

%\addlegendentry{SIC col}

\addplot+[smooth,color=red,solid, thick, every mark/.append style={solid} ,mark=o,
y filter/.code={\pgfmathparse{\pgfmathresult-0}\pgfmathresult}]
  table[row sep=crcr]{%
-15	0.162846679687500	\\
-10	0.121090429687500\\
-5	0.090581738281250	\\
0	0.065611621093750\\
5	0.049472558593750	\\
10	0.036302636718750	\\
15	0.027328320312500	\\
20	0.020413183593750\\
25	0.016409472656250	\\
30	0.013125878906250	\\
35	0.011013964843750	\\
40	0.009405175781250	\\
};
\addplot[smooth,color=black, thick, every mark/.append style={solid},
y filter/.code={\pgfmathparse{\pgfmathresult-0}\pgfmathresult}]
  table[row sep=crcr]{%
	-1 -2\\
};\label{P33}

\addplot[smooth,color=black,thick,dashed, every mark/.append style={solid}, 
y filter/.code={\pgfmathparse{\pgfmathresult-0}\pgfmathresult}]
  table[row sep=crcr]{%
	-1 -2\\
};\label{P34}

 \addplot[smooth,color=black,thick,dotted, every mark/.append style={dotted}, 
y filter/.code={\pgfmathparse{\pgfmathresult-0}\pgfmathresult}]
  table[row sep=crcr]{%
	-1 -1\\
};\label{P35}
\node [draw,fill=white, fill opacity=0.6,draw opacity=1,
text opacity =1,at ={(6,0.011)}, anchor= north west,draw=black,fill=white,font=\scriptsize]  {
\setlength{\tabcolsep}{0.5mm}
\renewcommand{\arraystretch}{.8}
\begin{tabular}{l}
 \ref{P35}{\hspace{0.15cm} Standard}\\
\ref{P34}{\hspace{0.15cm} LLR-Censoring}\\
\ref{P33}{\hspace{0.15cm} LLR-Combining}\\
\end{tabular}
};
 \end{axis}

%%
%% LIST
\begin{axis}[%
name=IF2,
    at={($(IF1.east)+(38,0em)$)},
		anchor= west,
%width=0.85\columnwidth,%5.63489583333333in,
width  = 0.7\columnwidth,%5.63489583333333in,
%height = 0.3\columnwidth,%4.16838541666667in,
height = 0.38\columnwidth,%4.16838541666667in,
scale only axis,
xmin  = -15,
xmax  = 40,
xlabel= {SNR [dB]},
xmajorgrids,
ymin=0.002,
ymax=0.5,
ymode=log,
ylabel={BER},
ymajorgrids,
legend entries={RMF,
					MMSE,
				MMSE-PIC,	
				%MMSE based comparator network,				
				},
legend style={fill=white, fill opacity=0.6, draw opacity=1,
text opacity =1,at={(0.02,0.03)}, anchor= south west,draw=black,fill=white,legend cell align=left,font=\scriptsize}
]

\addlegendimage{smooth,color=black,solid, thick, mark=x,
y filter/.code={\pgfmathparse{\pgfmathresult-0}\pgfmathresult}}
\addlegendimage{smooth,color=blue,solid, thick, mark=square,
y filter/.code={\pgfmathparse{\pgfmathresult-0}\pgfmathresult}}
\addlegendimage{smooth,color=red,solid, thick, mark=o,
y filter/.code={\pgfmathparse{\pgfmathresult-0}\pgfmathresult}}
\addlegendimage{smooth,color=magenta,solid, thick, mark=o,
y filter/.code={\pgfmathparse{\pgfmathresult-0}\pgfmathresult}}
\addlegendimage{smooth,color=green,solid, thick, mark=diamond,
y filter/.code={\pgfmathparse{\pgfmathresult-0}\pgfmathresult}}
% Add a (b) below x-axis
% Add a (c) below x-axis
\pgfplotsset{
    every axis/.append style={
        extra description/.code={
            \node at (0.5,-0.25) {(b)};
        },
    },
}

%\node at (axis cs:-5,1.3884) [anchor=south west] {$2 \cdot 0.6942$};

%\node at (axis cs:-5,1.3884) [anchor=north west] {max entropy rate $d=1$};
\addplot+[smooth,color=black,dotted,thick, every mark/.append style={solid} ,mark=x,
y filter/.code={\pgfmathparse{\pgfmathresult-0}\pgfmathresult}]
  table[row sep=crcr]{%
-15	0.374312972005209	\\
-10	0.350188151041666\\
-5	0.334508658854167	\\
0	0.322289322916667\\
5	0.314356591796875	\\
10	0.308103068033854\\
15	0.303589119466146	\\
20	0.301350008138021\\
25	0.299633235677084	\\
30	0.298243888346354	\\
35	0.297733349609375	\\
40	0.297426481119792	\\
%25	0	\\
};
%\addlegendentry{MMSE col}

\addplot+[smooth,color=blue,dotted, thick, every mark/.append style={solid} ,mark=square,
y filter/.code={\pgfmathparse{\pgfmathresult-0}\pgfmathresult}]
  table[row sep=crcr]{%
-15	0.362007714843750	\\
-10	0.332065722656250\\
-5	0.309230957031250	\\
0	0.287510758463541\\
5	0.269558398437500	\\
10	0.252333748372396\\
15	0.237404890950520	\\
20	0.224721752929688\\
25	0.213659008789063	\\
30	0.204875065104167\\
35	0.197552954101562	\\
40	0.191521704101563	\\
};

%\addlegendentry{SIC col}

\addplot+[smooth,color=red,dotted, thick, every mark/.append style={solid} ,mark=o,
y filter/.code={\pgfmathparse{\pgfmathresult-0}\pgfmathresult}]
  table[row sep=crcr]{%
-15	0.355763370768229	\\
-10	0.322455664062500\\
-5	0.294920930989584	\\
0	0.266942789713541\\
5	0.240917757161459	\\
10	0.215052596028646	\\
15	0.192108219401042	\\
20	0.172164770507813\\
25	0.155478369140625	\\
30	0.139886523437500	\\
35	0.128130851236979	\\
40	0.118438574218750	\\
};

%\node [draw,fill=white,font=\tiny,anchor= north east] at (axis cs: 28, 0.5 ) { $M_{\text{Tx}} = M_{\text{Rx}}= 3$ };

%% IMPERFECT CSI
\addplot+[smooth,color=black,dashed,thick, every mark/.append style={solid} ,mark=x,
y filter/.code={\pgfmathparse{\pgfmathresult-0}\pgfmathresult}]
  table[row sep=crcr]{%
-15	0.239232031250000	\\
-10	0.210764062500000\\
-5	0.195243066406250	\\
0	0.187946191406250\\
5	0.184578417968750	\\
10	0.182530859375000\\
15	0.181635546875000	\\
20	0.182250488281250\\
25	0.182270898437500	\\
30	0.181513476562500	\\
35	0.180764257812500\\
40	0.181268066406250	\\
%25	0	\\
};
%\addlegendentry{MMSE col}

\addplot+[smooth,color=blue,dashed, thick, every mark/.append style={solid} ,mark=square,
y filter/.code={\pgfmathparse{\pgfmathresult-0}\pgfmathresult}]
  table[row sep=crcr]{%
-15	0.196572656250000	\\
-10	0.155413183593750\\
-5	0.126651367187500	\\
0	0.105066992187500\\
5	0.088351074218750	\\
10	0.072676660156250\\
15	0.063320214843750	\\
20	0.053104785156250\\
25	0.047124316406250	\\
30	0.041366894531250\\
35	0.037770019531250	\\
40	0.034865332031250	\\
};

%\addlegendentry{SIC col}

\addplot+[smooth,color=red,dashed, thick, every mark/.append style={solid} ,mark=o,
y filter/.code={\pgfmathparse{\pgfmathresult-0}\pgfmathresult}]
  table[row sep=crcr]{%
-15	0.188163085937500	\\
-10	0.144539550781250\\
-5	0.112814746093750	\\
0	0.087320507812500\\
5	0.069278027343750	\\
10	0.053422656250000	\\
15	0.043974121093750	\\
20	0.034550000000000\\
25	0.029682421875000	\\
30	0.024958593750000	\\
35	0.021893359375000	\\
40	0.019589257812500	\\
};
\addplot+[smooth,color=black,solid,thick, every mark/.append style={solid} ,mark=x,
y filter/.code={\pgfmathparse{\pgfmathresult-0}\pgfmathresult}]
  table[row sep=crcr]{%
-15	0.224790136718750	\\
-10	0.196068066406250\\
-5	0.180007519531250	\\
0	0.171773437500000\\
5	0.168293945312500	\\
10	0.165648535156250\\
15	0.163943164062500	\\
20	0.164057617187500\\
25	0.164264941406250	\\
30	0.164307421875000	\\
35	0.163024414062500	\\
40	0.163983496093750	\\
%25	0	\\
};
%\addlegendentry{MMSE col}

\addplot+[smooth,color=blue,solid, thick, every mark/.append style={solid} ,mark=square,
y filter/.code={\pgfmathparse{\pgfmathresult-0}\pgfmathresult}]
  table[row sep=crcr]{%
-15	0.183333398437500	\\
-10	0.142945117187500\\
-5	0.114198339843750	\\
0	0.092925585937500\\
5	0.076601855468750	\\
10	0.062050781250000\\
15	0.053222460937500	\\
20	0.043867480468750\\
25	0.038452734375000	\\
30	0.033571875000000\\
35	0.030166992187500	\\
40	0.027992773437500	\\
};

%\addlegendentry{SIC col}

\addplot+[smooth,color=red,solid, thick, every mark/.append style={solid} ,mark=o,
y filter/.code={\pgfmathparse{\pgfmathresult-0}\pgfmathresult}]
  table[row sep=crcr]{%
-15	0.175355468750000	\\
-10	0.132869628906250\\
-5	0.101450097656250	\\
0	0.076593945312500\\
5	0.059347070312500	\\
10	0.044971484375000	\\
15	0.035690917968750	\\
20	0.027519335937500\\
25	0.023275292968750	\\
30	0.019406835937500	\\
35	0.016711035156250	\\
40	0.014750878906250	\\
};
\addplot[smooth,color=black, thick, every mark/.append style={solid},
y filter/.code={\pgfmathparse{\pgfmathresult-0}\pgfmathresult}]
  table[row sep=crcr]{%
	-1 -2\\
};\label{P33}

\addplot[smooth,color=black,thick,dashed, every mark/.append style={solid}, 
y filter/.code={\pgfmathparse{\pgfmathresult-0}\pgfmathresult}]
  table[row sep=crcr]{%
	-1 -2\\
};\label{P34}

 \addplot[smooth,color=black,thick,dotted, every mark/.append style={dotted}, 
y filter/.code={\pgfmathparse{\pgfmathresult-0}\pgfmathresult}]
  table[row sep=crcr]{%
	-1 -1\\
};\label{P35}
\node [draw,fill=white, fill opacity=0.6,draw opacity=1,
text opacity =1,at ={(6,0.011)}, anchor= north west,draw=black,fill=white,font=\scriptsize]  {
\setlength{\tabcolsep}{0.5mm}
\renewcommand{\arraystretch}{.8}
\begin{tabular}{l}
 \ref{P35}{\hspace{0.15cm} Standard}\\
\ref{P34}{\hspace{0.15cm} LLR-Censoring}\\
\ref{P33}{\hspace{0.15cm} LLR-Combining}\\
\end{tabular}
};

\end{axis}
\end{tikzpicture}%

%% file: IDDITERVARY.tex
\definecolor{mycolor1}{rgb}{0.00000,1.00000,1.00000}%
\definecolor{mycolor2}{rgb}{1.00000,0.00000,1.00000}%
\definecolor{mycolor3}{rgb}{0.83,0.69,0.22}%

\pgfplotsset{every axis label/.append style={font=\scriptsize
},
every tick label/.append style={font=\scriptsize
}
}

\begin{tikzpicture}[font=\scriptsize
] 
\begin{axis}[%
name=IF1,
%ymode=log,
width  = 0.7\columnwidth,%5.63489583333333in,
%height = 0.3\columnwidth,%4.16838541666667in,
height = 0.34\columnwidth,%4.16838541666667in,
scale only axis,
xmin  = -15,
xmax  = 40,
xlabel= {SNR [dB]},
xmajorgrids,
ymin=0.002,
ymax=0.5,
ymode=log,
ylabel={BER},
ymajorgrids,
legend entries={\text{IDD = 1},
					\text{IDD = 2},
				\text{IDD = 3},	
				%MMSE based comparator network,				
				},
legend style={fill=white, fill opacity=0.6, draw opacity=1,
text opacity =1,at={(0.02,0.03)}, anchor= south west,draw=black,fill=white,legend cell align=left,font=\scriptsize}
]

\addlegendimage{smooth,color=magenta,solid, thick, mark=x,
y filter/.code={\pgfmathparse{\pgfmathresult-0}\pgfmathresult}}
\addlegendimage{smooth,color=cyan,solid, thick, mark=square,
y filter/.code={\pgfmathparse{\pgfmathresult-0}\pgfmathresult}}
\addlegendimage{smooth,color=orange,solid, thick, mark=o,
y filter/.code={\pgfmathparse{\pgfmathresult-0}\pgfmathresult}}
\addlegendimage{smooth,color=magenta,solid, thick, mark=o,
y filter/.code={\pgfmathparse{\pgfmathresult-0}\pgfmathresult}}
\addlegendimage{smooth,color=green,solid, thick, mark=diamond,
y filter/.code={\pgfmathparse{\pgfmathresult-0}\pgfmathresult}}
% Add a (b) below x-axis
\pgfplotsset{
    every axis/.append style={
        extra description/.code={
            \node at (0.5,-0.29) {(c)};
        },
    },
}

% \draw (31,0.18) ellipse (0.2cm and 0.6cm);
% \draw[dspconn]    (30,0.25) -- (20,0.3) ;
% \draw (12,0.4) node [anchor=north west][inner sep=0.75pt]  [font=\footnotesize]  {\text{All-APs}};

% \draw (35,0.05) ellipse (0.2cm and 0.9cm);
% \draw[dspconn]   (30,0.04) -- (10,0.055) ;
% \draw (4,0.07) node [anchor=north west][inner sep=0.75pt]  [font=\footnotesize]  {\text{APs-Sel}};
\addplot+[smooth,color=magenta,dotted,thick, every mark/.append style={solid} ,mark=x,
y filter/.code={\pgfmathparse{\pgfmathresult-0}\pgfmathresult}]
  table[row sep=crcr]{%
-15	0.359961791992188	\\
-10	0.332586669921875\\
-5	0.309347656250000	\\
0	0.287295410156250\\
5	0.268310058593750	\\
10	0.253827514648438\\
15	0.237430908203125	\\
20	0.224513549804688\\
25	0.213888183593750	\\
30	0.203719726562500\\
35	0.197688110351563	\\
40	0.193306884765625	\\
%25	0	\\
};
%\addlegendentry{MMSE col}

\addplot+[smooth,color=cyan,dotted, thick, every mark/.append style={solid} ,mark=square,
y filter/.code={\pgfmathparse{\pgfmathresult-0}\pgfmathresult}]
  table[row sep=crcr]{%
-15	0.353482177734375	\\
-10	0.322436279296875\\
-5	0.294931884765625	\\
0	0.264911376953125\\
5	0.237693847656250	\\
10	0.215501098632813	\\
15	0.191346435546875	\\
20	0.170217895507813\\
25	0.154580322265625	\\
30	0.138890014648438	\\
35	0.127731079101563	\\
40	0.120689331054688	\\
};

%\addlegendentry{SIC col}

\addplot+[smooth,color=orange,dotted, thick, every mark/.append style={solid} ,mark=o,
y filter/.code={\pgfmathparse{\pgfmathresult-0}\pgfmathresult}]
  table[row sep=crcr]{%
-15	0.353196411132813	\\
-10	0.321676513671875\\
-5	0.293425537109375	\\
0	0.262989990234375\\
5	0.234918457031250	\\
10	0.212100830078125	\\
15	0.186718139648438	\\
20	0.164589843750000\\
25	0.148190307617188	\\
30	0.130995849609375	\\
35	0.118631469726563	\\
40	0.109959838867188	\\
};
%% IMPERFECT CSI

\addplot+[smooth,color=magenta,dashed,thick, every mark/.append style={solid} ,mark=x,
y filter/.code={\pgfmathparse{\pgfmathresult-0}\pgfmathresult}]
  table[row sep=crcr]{%
-15	0.184824511718750	\\
-10	0.144231250000000\\
-5	0.116271875000000	\\
0	0.094084863281250\\
5	0.078165917968750	\\
10	0.063386718750000\\
15	0.053987500000000	\\
20	0.044621582031250\\
25	0.038811328125000	\\
30	0.033607910156250\\
35	0.030712304687500	\\
40	0.027699414062500	\\
%25	0	\\
};
%\addlegendentry{MMSE col}

\addplot+[smooth,color=cyan,dashed, thick, every mark/.append style={solid} ,mark=square,
y filter/.code={\pgfmathparse{\pgfmathresult-0}\pgfmathresult}]
  table[row sep=crcr]{%
-15	0.176170507812500	\\
-10	0.133085449218750\\
-5	0.102417675781250	\\
0	0.076673535156250\\
5	0.059537011718750	\\
10	0.044890234375000	\\
15	0.035723632812500	\\
20	0.027275976562500\\
25	0.022764843750000	\\
30	0.018699218750000	\\
35	0.016093945312500	\\
40	0.014187597656250	\\
};

%\addlegendentry{SIC col}

\addplot+[smooth,color=orange,dashed, thick, every mark/.append style={solid} ,mark=o,
y filter/.code={\pgfmathparse{\pgfmathresult-0}\pgfmathresult}]
  table[row sep=crcr]{%
-15	0.175836523437500	\\
-10	0.132708789062500\\
-5	0.101880371093750	\\
0	0.076007910156250\\
5	0.058763769531250	\\
10	0.043985937500000	\\
15	0.034729882812500	\\
20	0.026390039062500\\
25	0.021686328125000	\\
30	0.017826757812500	\\
35	0.015095019531250	\\
40	0.013278222656250	\\
};

%\addlegendentry{MF SIC col}

\addplot+[smooth,color=magenta, solid,thick, every mark/.append style={solid} ,mark=x,
y filter/.code={\pgfmathparse{\pgfmathresult-0}\pgfmathresult}]
  table[row sep=crcr]{%
-15	0.170993652343750	\\
-10	0.131309960937500\\
-5	0.103304687500000	\\
0	0.081506835937500\\
5	0.066121191406250	\\
10	0.052459960937500\\
15	0.043685546875000	\\
20	0.035437402343750\\
25	0.030042871093750	\\
30	0.025676367187500\\
35	0.022973632812500	\\
40	0.020736621093750	\\
%25	0	\\
};
%\addlegendentry{MMSE col}

\addplot+[smooth,color=cyan, solid, thick, every mark/.append style={solid} ,mark=square,
y filter/.code={\pgfmathparse{\pgfmathresult-0}\pgfmathresult}]
  table[row sep=crcr]{%
-15	0.162846679687500	\\
-10	0.121090429687500\\
-5	0.090581738281250	\\
0	0.065611621093750\\
5	0.049472558593750	\\
10	0.036302636718750	\\
15	0.027328320312500	\\
20	0.020413183593750\\
25	0.016409472656250	\\
30	0.013125878906250	\\
35	0.011013964843750	\\
40	0.009405175781250	\\
};

%\addlegendentry{SIC col}

\addplot+[smooth,color=orange,solid, thick, every mark/.append style={solid} ,mark=o,
y filter/.code={\pgfmathparse{\pgfmathresult-0}\pgfmathresult}]
  table[row sep=crcr]{%
-15	0.162582519531250	\\
-10	0.120735058593750\\
-5	0.090130078125000	\\
0	0.065077929687500\\
5	0.048834960937500	\\
10	0.035562304687500	\\
15	0.026432714843750	\\
20	0.019675878906250\\
25	0.015541015625000	\\
30	0.012438085937500	\\
35	0.010306347656250	\\
40	0.008769238281250	\\
};

\addplot[smooth,color=black, thick, every mark/.append style={solid},
y filter/.code={\pgfmathparse{\pgfmathresult-0}\pgfmathresult}]
  table[row sep=crcr]{%
	-1 -2\\
};\label{P33}

\addplot[smooth,color=black,thick,dashed, every mark/.append style={solid}, 
y filter/.code={\pgfmathparse{\pgfmathresult-0}\pgfmathresult}]
  table[row sep=crcr]{%
	-1 -2\\
};\label{P34}

 \addplot[smooth,color=black,thick,dotted, every mark/.append style={dotted}, 
y filter/.code={\pgfmathparse{\pgfmathresult-0}\pgfmathresult}]
  table[row sep=crcr]{%
	-1 -1\\
};\label{P35}
\node [draw,fill=white, fill opacity=0.6,draw opacity=1,
text opacity =1,at ={(3,0.013)}, anchor= north west,draw=black,fill=white,font=\scriptsize]  {
\setlength{\tabcolsep}{0.5mm}
\renewcommand{\arraystretch}{.8}
\begin{tabular}{l}
 \ref{P35}{\hspace{0.15cm} Standard}\\
\ref{P34}{\hspace{0.15cm} LLR-Censoring}\\
\ref{P33}{\hspace{0.15cm} LLR-Combining}\\
\end{tabular}
};
 \end{axis}

%%
%% LIST
\begin{axis}[%
name=IF2,
    at={($(IF1.east)+(38,0em)$)},
		anchor= west,
%width=0.85\columnwidth,%5.63489583333333in,
width  = 0.7\columnwidth,%5.63489583333333in,
%height = 0.3\columnwidth,%4.16838541666667in,
height = 0.34\columnwidth,%4.16838541666667in,
scale only axis,
xmin  = -15,
xmax  = 40,
xlabel= {SNR [dB]},
xmajorgrids,
ymin=0.002,
ymax=0.5,
ymode=log,
ylabel={BER},
ymajorgrids,
legend entries={\text{IDD = 1},
					\text{IDD = 2},
				\text{IDD = 3},	
				%MMSE based comparator network,				
				},
legend style={fill=white, fill opacity=0.6, draw opacity=1,
text opacity =1,at={(0.02,0.03)}, anchor= south west,draw=black,fill=white,legend cell align=left,font=\scriptsize}
]

\addlegendimage{smooth,color=magenta,solid, thick, mark=x,
y filter/.code={\pgfmathparse{\pgfmathresult-0}\pgfmathresult}}
\addlegendimage{smooth,color=cyan,solid, thick, mark=square,
y filter/.code={\pgfmathparse{\pgfmathresult-0}\pgfmathresult}}
\addlegendimage{smooth,color=orange,solid, thick, mark=o,
y filter/.code={\pgfmathparse{\pgfmathresult-0}\pgfmathresult}}
\addlegendimage{smooth,color=magenta,solid, thick, mark=o,
y filter/.code={\pgfmathparse{\pgfmathresult-0}\pgfmathresult}}
\addlegendimage{smooth,color=green,solid, thick, mark=diamond,
y filter/.code={\pgfmathparse{\pgfmathresult-0}\pgfmathresult}}
% Add a (b) below x-axis
% Add a (c) below x-axis
\pgfplotsset{
    every axis/.append style={
        extra description/.code={
            \node at (0.5,-0.29) {(d)};
        },
    },
}

%\node at (axis cs:-5,1.3884) [anchor=south west] {$2 \cdot 0.6942$};

%\node at (axis cs:-5,1.3884) [anchor=north west] {max entropy rate $d=1$};

%\node [draw,fill=white,font=\tiny,anchor= north east] at (axis cs: 28, 0.5 ) { $M_{\text{Tx}} = M_{\text{Rx}}= 3$ };
\addplot+[smooth,color=magenta,dotted,thick, every mark/.append style={solid} ,mark=x,
y filter/.code={\pgfmathparse{\pgfmathresult-0}\pgfmathresult}]
  table[row sep=crcr]{%
-15	0.362007714843750	\\
-10	0.332065722656250\\
-5	0.309230957031250	\\
0	0.287510758463541\\
5	0.269558398437500	\\
10	0.252333748372396\\
15	0.237404890950520	\\
20	0.224721752929688\\
25	0.213659008789063	\\
30	0.204875065104167\\
35	0.197552954101562	\\
40	0.191521704101563	\\
%25	0	\\
};
%\addlegendentry{MMSE col}

\addplot+[smooth,color=cyan,dotted, thick, every mark/.append style={solid} ,mark=square,
y filter/.code={\pgfmathparse{\pgfmathresult-0}\pgfmathresult}]
  table[row sep=crcr]{%
-15	0.355763370768229	\\
-10	0.322455664062500\\
-5	0.294920930989584	\\
0	0.266942789713541\\
5	0.240917757161459	\\
10	0.215052596028646	\\
15	0.192108219401042	\\
20	0.172164770507813\\
25	0.155478369140625	\\
30	0.139886523437500	\\
35	0.128130851236979	\\
40	0.118438574218750	\\
};

%\addlegendentry{SIC col}

\addplot+[smooth,color=orange,dotted, thick, every mark/.append style={solid} ,mark=o,
y filter/.code={\pgfmathparse{\pgfmathresult-0}\pgfmathresult}]
  table[row sep=crcr]{%
-15	0.355443766276042	\\
-10	0.321694783528646\\
-5	0.293860408528646	\\
0	0.265532918294271\\
5	0.238628515625000	\\
10	0.211745751953125	\\
15	0.187696313476562	\\
20	0.166234277343750\\
25	0.148066357421875	\\
30	0.131974527994792	\\
35	0.119026529947917	\\
40	0.109036108398438	\\
};

%% IMPERFECT CSI

%\addlegendentry{MF SIC col}

\addplot+[smooth,color=magenta,dashed,thick, every mark/.append style={solid} ,mark=x,
y filter/.code={\pgfmathparse{\pgfmathresult-0}\pgfmathresult}]
  table[row sep=crcr]{%
-15	0.197228808593750	\\
-10	0.154493554687500\\
-5	0.126934960937500	\\
0	0.104546289062500\\
5	0.087272265625000	\\
10	0.074140722656250\\
15	0.062636914062500	\\
20	0.053996777343750\\
25	0.046487011718750	\\
30	0.041857519531250\\
35	0.037644824218750	\\
40	0.034690722656250	\\
%25	0	\\
};
%\addlegendentry{MMSE col}

\addplot+[smooth,color=cyan, dashed, thick, every mark/.append style={solid} ,mark=square,
y filter/.code={\pgfmathparse{\pgfmathresult-0}\pgfmathresult}]
  table[row sep=crcr]{%
-15	0.188565332031250	\\
-10	0.143277343750000\\
-5	0.113058105468750	\\
0	0.087408203125000\\
5	0.068244921875000	\\
10	0.054389062500000	\\
15	0.043270019531250	\\
20	0.035610839843750\\
25	0.028909179687500	\\
30	0.025132812500000	\\
35	0.022535253906250	\\
40	0.019344335937500	\\
};

%\addlegendentry{SIC col}

\addplot+[smooth,color=orange,dashed, thick, every mark/.append style={solid} ,mark=o,
y filter/.code={\pgfmathparse{\pgfmathresult-0}\pgfmathresult}]
  table[row sep=crcr]{%
-15	0.188327148437500	\\
-10	0.142915136718750\\
-5	0.112602539062500	\\
0	0.086815234375000\\
5	0.067481640625000	\\
10	0.053441503906250	\\
15	0.042204687500000	\\
20	0.034591992187500\\
25	0.027792675781250	\\
30	0.024110839843750	\\
35	0.021370800781250	\\
40	0.018199414062500	\\
};
\addplot+[smooth,color=magenta,solid,thick, every mark/.append style={solid} ,mark=x,
y filter/.code={\pgfmathparse{\pgfmathresult-0}\pgfmathresult}]
  table[row sep=crcr]{%
-15	0.184020410156250	\\
-10	0.141735546875000\\
-5	0.114397656250000	\\
0	0.092495117187500\\
5	0.075640625000000	\\
10	0.063644238281250\\
15	0.052670703125000	\\
20	0.044547558593750\\
25	0.038224316406250	\\
30	0.033710156250000\\
35	0.030337597656250	\\
40	0.027490039062500	\\
%25	0	\\
};
%\addlegendentry{MMSE col}

\addplot+[smooth,color=cyan,solid, thick, every mark/.append style={solid} ,mark=square,
y filter/.code={\pgfmathparse{\pgfmathresult-0}\pgfmathresult}]
  table[row sep=crcr]{%
-15	0.175944628906250	\\
-10	0.131374902343750\\
-5	0.101579199218750	\\
0	0.076705957031250\\
5	0.058560351562500	\\
10	0.045868359375000\\
15	0.035566113281250	\\
20	0.028716015625000\\
25	0.022716113281250	\\
30	0.019504296875000	\\
35	0.017132910156250	\\
40	0.014337988281250	\\
};

%\addlegendentry{SIC col}

\addplot+[smooth,color=orange,solid, thick, every mark/.append style={solid} ,mark=o,
y filter/.code={\pgfmathparse{\pgfmathresult-0}\pgfmathresult}]
  table[row sep=crcr]{%
-15	0.175737207031250	\\
-10	0.131052929687500\\
-5	0.101182421875000	\\
0	0.076183007812500\\
5	0.057866894531250	\\
10	0.045107714843750	\\
15	0.034674414062500	\\
20	0.027804296875000\\
25	0.021764746093750	\\
30	0.018692675781250	\\
35	0.016238378906250	\\
40	0.013506542968750	\\
};
\addplot[smooth,color=black, thick, every mark/.append style={solid},
y filter/.code={\pgfmathparse{\pgfmathresult-0}\pgfmathresult}]
  table[row sep=crcr]{%
	-1 -2\\
};\label{P33}

\addplot[smooth,color=black,thick,dashed, every mark/.append style={solid}, 
y filter/.code={\pgfmathparse{\pgfmathresult-0}\pgfmathresult}]
  table[row sep=crcr]{%
	-1 -2\\
};\label{P34}

 \addplot[smooth,color=black,thick,dotted, every mark/.append style={dotted}, 
y filter/.code={\pgfmathparse{\pgfmathresult-0}\pgfmathresult}]
  table[row sep=crcr]{%
	-1 -1\\
};\label{P35}
\node [draw,fill=white, fill opacity=0.6,draw opacity=1,
text opacity =1,at ={(3,0.015)}, anchor= north west,draw=black,fill=white,font=\scriptsize]  {
\setlength{\tabcolsep}{0.5mm}
\renewcommand{\arraystretch}{.8}
\begin{tabular}{l}
 \ref{P35}{\hspace{0.15cm} Standard}\\
\ref{P34}{\hspace{0.15cm} LLR-Censoring}\\
\ref{P33}{\hspace{0.15cm} LLR-Combining}\\
\end{tabular}
};
\end{axis}
\end{tikzpicture}%